# A Route to Unusually Broadband Absorption Spanning from Visible to Mid-Infrared

Majid Aalizadeh[1,2,*], Amin Khavasi[3], Andriy E. Serebryannikov[4,5], Guy A. E. Vandenbosch[4], and Ekmel Ozbay[1,2,6,7]

[1] Department of Electrical and Electronics Engineering, Bilkent University, Ankara 06800, Turkey
[2] Nanotechnology Research Center (NANOTAM), Bilkent University, Ankara 06800, Turkey
[3] Electrical Engineering Department, Sharif University of Technology, Tehran, 11155-4363, Iran
[4] ESAT-TELEMIC, Katholieke Universiteit Leuven, 3000 Leuven, Belgium
[5] Faculty of Physics, Adam Mickiewicz University, 61-614 Poznan, Poland
[6] National Nanotechnology Research Center (UNAM), Bilkent University, Ankara 06800, Turkey
[7] Department of Physics, Bilkent University, Ankara 06800, Turkey
*majid.aalizadeh@bilkent.edu.tr

## Abstract

In this paper, a route to ultra-broadband absorption is suggested and demonstrated by a feasible design. The high absorption regime (absorption above 90%) for the suggested structure ranges from visible to mid-infrared (MIR), i.e. for the wavelength from 478 to 3,278 nm that yields an ultra-wide bandwidth of 2,800 nm. The structure consists of a top-layer-patterned metal-insulator-metal (MIM) configuration, into the insulator layer of which, an ultra-thin 5 nm layer of Manganese (Mn) is embedded. The MIM configuration represents a Ti-Al$_2$O$_3$-Ti tri-layer. It is shown that, without the ultra-thin layer of Mn, the absorption bandwidth is reduced to 274 nm. Therefore, adding only a 5 nm layer of Mn leads to a more than tenfold increase in the width of the absorption band. It is explained in detail that the physical mechanism contributing to this ultra-broadband result is a combination of plasmonic and non-plasmonic resonance modes, along with the appropriate optical properties of Mn. This structure has the relative bandwidth (RBW) of 149%, while only one step of lithography is required for its fabrication, so it is relatively simple to fabricate. This makes it rather promising for practical applications.

**Keywords:** localized surface plasmons, nanodisk array, impedance matching, guided-mode resonance

## 1. Introduction

Electromagnetic (EM) absorbers have gained a lot of attention due to their wide applications in various fields, including photovoltaics [1,2], sensing [3,4], photodetection [5,6], thermal imaging [7], thermal emission [8], shielding [9,10], etc. Metals are natural absorbers; however, they are required to be very thick to perfectly absorb the incoming EM wave's energy. This requirement for the thickness or volume of metals to be used as absorbers is in huge contradiction with the down-scaling trend of all optical and photonic devices. The concept of metamaterials has made it possible to overcome this contradiction and to come up with EM absorbers with a subwavelength thickness. Such metamaterial-based EM absorbers are reported in various frequency regimes that include microwave [11], terahertz [12,13], far-Infrared (FIR) [14], mid-Infrared (MIR) [15], near-infrared (NIR) [16], and visible [17,18].

Light absorption can be realized by trapping the light in an EM resonance mode and dissipating its power through the intrinsic loss of materials. The typical resonance modes used in absorbers include Localized Surface Plasmon (LSP) [19,20], Propagating Surface Plasmon (PSP) [21,22], and Fabry-Perot resonance [23,24] modes. Broadband metamaterial absorbers have all the earlier mentioned applications, except for sensing. This is obviously because of the fact that, to achieve a better sensing functionality, a narrower resonance is typically desired [25]. The main approaches for obtaining broadband absorption are either through broadening the resonance mode or by using the superposition of several adjacent resonances, or a combination of both. Broadening the resonances is equivalent to decreasing the quality-factor of the resonance mode. This is possible by the appropriate geometric design and through the use of metals with high loss.

It is also possible to obtain perfect absorption while not having any specific resonance mode. It can be realized by blocking both the transmission and reflection. The transmission can be blocked by using an optically thick metal layer as the most bottom layer of the structure, and the reflection can be suppressed by providing impedance matching with free-space [13,26]. Therefore, it is possible to have perfect absorption by impedance matching and without any resonance mode. This is mostly useful for absorbing at the wavelengths that are much larger than the dimensions of the array's unit-cell.

In one of the early works on broadband absorbers, Aydin et al. experimentally achieved 71 percent average absorption in the visible (400-700 nm) range [27]. It was obtained using nano-rods with a non-uniform (tapered) width, which led to the superposition of various LSP resonances each corresponding to a specific width. There are also some lithography-free (non-arrayed) broadband absorbers that are fabricated in our group by the use of dewetting mechanism [28-30]. Dewetting is a method in which ultra-thin layers are annealed at high temperatures to deform into nanoholes or nanoparticles, depending on the annealing recipe [31]. The nanoholes or nanoparticles with random sizes then lead to having the superposition of LSP modes each corresponding to a specific nanoparticle size. There is also another recent lithography-free work carried out in our group in which a broadband absorption was achieved by the superposition of LSP resonances of randomly oriented dielectric-metal core-shell nanowires [32]. Broadening the Fabry-Perot resonance mode has also been extensively studied in MIM cavities [33,34] and Metal-Insulator stacks [35,36].

One common absorber structure is the MIM structure, in which the top metal layer is patterned using lithography. In one of the recent works, Au-$SiO_2$-Ti configuration is used in the MIM structure, and the top Ti layer is patterned as a periodic array of nanodisks [16]. The high absorption of that structure is extended over the wavelength range of 900 to 1,825 nm with a 71% percent relative bandwidth that is defined as the ratio of the high absorption bandwidth to the center wavelength.

In this work, we show that, by embedding an ultra-thin 5 nm layer of Mn into the dielectric layer of a typical top-layer-patterned MIM structure (with a 2D array of nanodisks on the top), an ultra-broadband, nearly perfect absorption (absorption greater than 0.9) can be obtained over the wavelength range of 478 to 3,278 nm. The bandwidth and relative bandwidth of the proposed structure are 2,800 nm and 149%, respectively, and the absorption band extends over the visible, NIR, and MIR regions. This is, to our knowledge, the best result obtained for such absorbers with one-stage-lithography. The absorption band of the MIM structure without the ultra-thin Mn layer covers the two adjacent ranges of 478 to 526 and 610 to 836 nm that is equivalent to 274 nm of total bandwidth. Therefore, adding a thin layer of Mn, leads to a 10.12 fold increase in the bandwidth, without adding much complexity or lithography stage to the fabrication process. A step-by-step dimension and material optimization procedure is presented. We investigate different types of metals in order to find the most appropriate ones for an ultra-wideband absorber. We also consider the effect of the dimensions of the structure on the absorption bandwidth. Then, the physical mechanisms that lead to strong absorption in different wavelength ranges are discussed. In particular, we discuss the importance of the 5 nm Mn layer for the drastic increase of the bandwidth of the structure. This is at a time when Mn has not been given the attention it deserves for its use in broadband absorbers. Mn was used for the first and, up to now, the only time in a broadband absorber structure that was recently fabricated in our group, and it led to a very promising result compared to other metals [34]. Moreover, we demonstrate the high fabrication tolerance of the proposed structure. Finally, we explore the effect of polarization and incidence angle. Numerical results are obtained using the finite difference time domain (FDTD) method by Lumerical FDTD software [37], and the results are verified by regenerating them using the finite integration technique (FIT) by CST Microwave Studio [38].

It is very important to mention that, because of having one-level lithography, the large-area fabrication of the proposed structure can be realized by the use of the standard nanoimprint lithography technique [39]. This is very beneficial for mass and high-throughput production purposes.

## 2. Design and optimization procedure

### 2.1 Proposed structure

The schematic of the proposed absorber is shown in Fig. 1. It consists of an MIM structure with the top and bottom metal layers being chosen as Ti, and $Al_2O_3$ being used as the dielectric layer, and with the top Ti layer being patterned to obtain a periodic array of nanodisks, while an ultra-thin, 5 nm layer of Mn is embedded into the dielectric layer. As shown in Figs. 1(a-c), the period, height of nanodisks, radius of nanodisks, thickness of the Mn layer, the distance between the bottom metal and the top nanodisks, and the distance between the bottom metal and the Mn layer, are denoted by p, h, r, $t_{Mn}$, $d_i$ and $d_{Mn}$, respectively. The bottom metal layer is expected to be sufficiently thick to block transmission. The perspective description of the proposed absorber is shown in Fig. 1(d). In this view, the 5 nm Mn layer can be considered as a perturbation that is introduced to the patterned MIM configuration of Ti-$Al_2O_3$-Ti. This is while, as it will be elevated in the forthcoming sections, the effect of this ultra-thin Mn layer is drastically significant in the enhancement and broadening of the nearly perfect absorption band. The optimized dimensions for the structure, according to Fig. 1, will be obtained in the next sections. They are equal to 480, 140, 180, 165, 5, and 135 nm for p, r, h, $d_i$, $t_{Mn}$, and $d_{Mn}$, respectively. The thickness of the bottom metal layer is conservatively chosen to be 200 nm to ensure nearly zero transmission. It should be mentioned that this choice of metals is the result of simulating all the possible combinations of several metals (which are conventionally used in broadband absorbers) in the proposed structure, and performing a thorough study on the results. These metals include Chromium (Cr), Tungsten (W), Platinum (Pt), and Iron (Fe), in addition to Ti and Mn.

### 2.2 Material optimizations

In order to demonstrate the effect of the choice of materials on the performance of the device and to optimize the structure in terms of materials, the absorption spectra have been simulated. The results are shown in Figs. 2(a), (b), (c), and (d) for the case of only changing the dielectric, bottom metal, top nanodisks, and 5 nm ultra-thin middle layer, respectively. All of the other materials and dimensions of the structure were kept unchanged. The thin black dashed line in all figures corresponds to 90 percent absorption. The refractive index data of all the metals except Mn, along with $Al_2O_3$ and $SiO_2$ are taken from [40], and the refractive index data of Mn, $HfO_2$, and $TiO_2$ are taken from [41], [42], and [43], respectively. Our purpose here is to choose the configuration that gives the widest bandwidth, i.e. to maximize the wavelength range in which the absorption is above 90 percent. The absorption (A) in general case can be obtained given the values of reflection (R) and transmission (T), using the formula A=1-R-T, where R=$|r|^2$ and T=$|t|^2$ with r and t being reflection and transmission coefficients, respectively. However, since the bottom metallic layer is sufficiently thick to suppress any transmission (200 nm), the absorption formula simplifies to A=1-R.

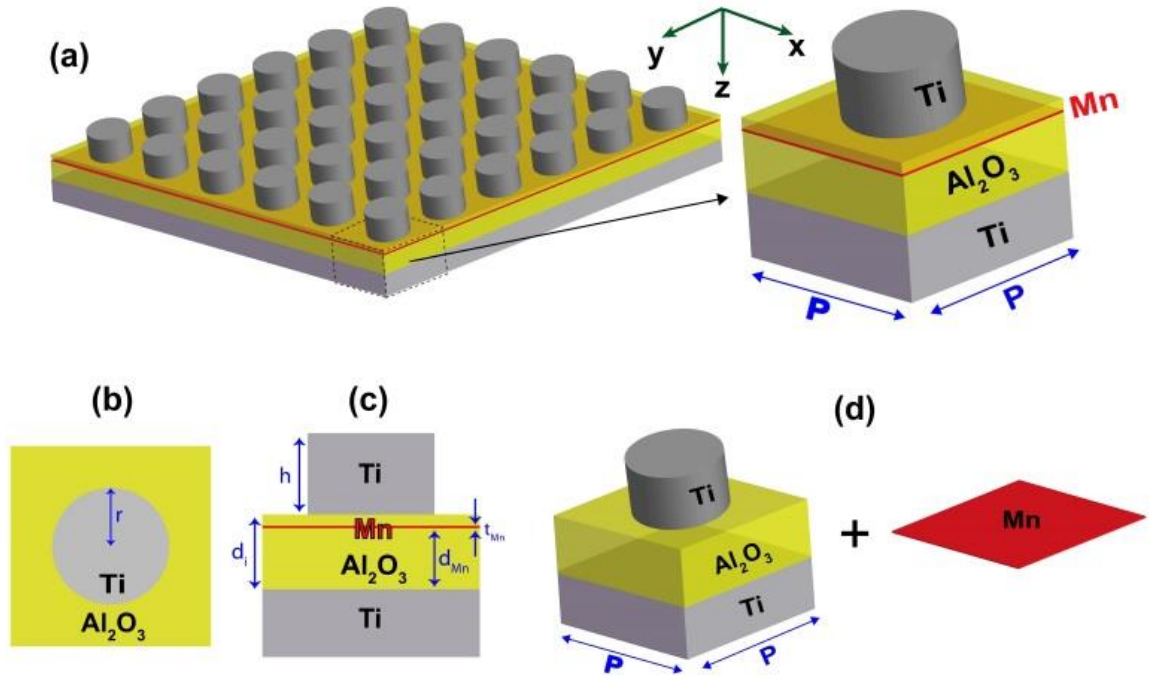

Fig. 1. (a) Left plot: 3D schematic of the absorber; one unit cell of the structure is bounded by dashed lines; right plot: a magnified view of one unit cell. (b) Top view, and (c) side view of the structure; and (d) the other image of the unit cell of the studied structure (perspective view): a patterned MIM configuration with top metal being made of periodic nanodisks, plus an ultra-thin Mn layer embedded into the insulator of the original configuration.

Let us first consider the effect of dielectric material on the absorption spectrum. The dielectrics chosen to be used in this investigation are $SiO_2$, $Al_2O_3$, $HfO_2$, and $TiO_2$, which are written from small to large refractive indices, respectively. It can be observed in Fig. 2(a) that by choosing dielectrics with higher refractive index, almost all the absorption peaks experience a red-shift. This is already expected to happen because of the nature of the resonance modes [44]. This figure demonstrates that the best bandwidth is obtained for the case of using $Al_2O_3$ as the dielectric material. For the case of changing the materials of the bottom metal and the top nanodisks, as shown in Figs. 2(b) and (c), respectively, it can be noticed that the absorption does not experience drastic changes. This shows that the absorption capability of the entire structure, when taking the bottom layer and top nanodisks into account, dominantly results from the geometric design rather than the choice of materials. However, about the middle metal layer, which is taken as 5 nm, it can be observed in Fig. 2(d) that the choice of metal drastically affects the absorption spectrum's shape and bandwidth at wavelengths larger than 800 nm. Therefore, this ultra-thin layer plays an important role in broadening the absorption spectrum in the NIR and MIR regions. The results shown in Figs. 2(a-d) justify that the choice of the materials in Fig. 1 provides the widest absorption band.

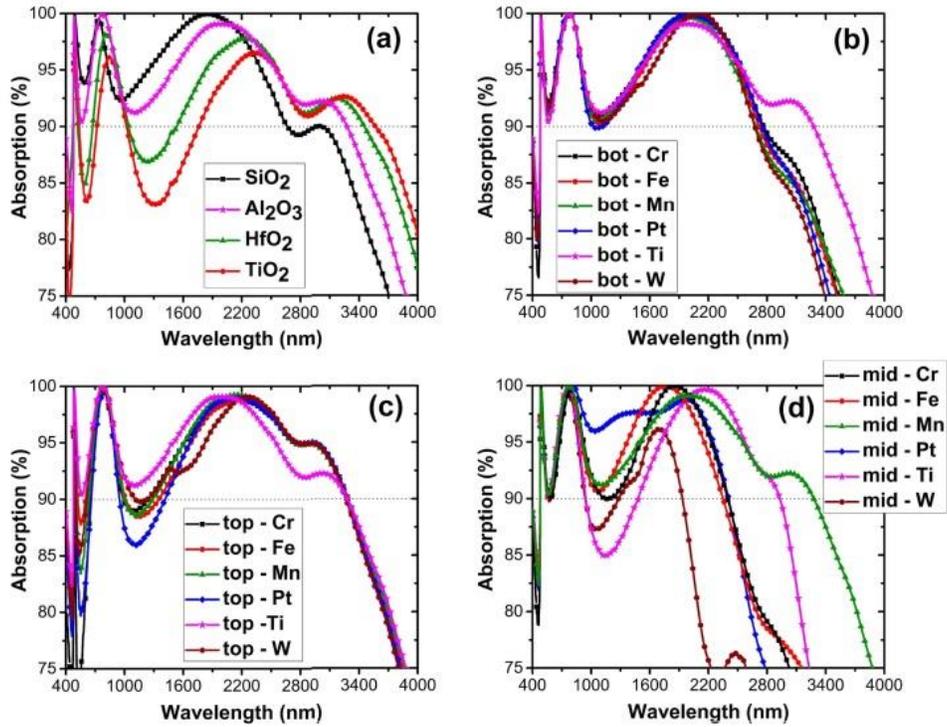

Fig. 2. Absorption spectra for different materials of the (a) dielectric layer, (b) bottom metal layer, (c) top metallic nanodisks, and (d) middle ultra-thin metal layer, while using the same materials for the remaining components as in Fig. 1. The dimensions of the structure in all cases are: p = 480 nm, r = 140 nm, h = 180 nm, $d_i$ = 165 nm, $t_{Mn}$ = 5 nm, and $d_{Mn}$ = 135 nm.

## 2.3 Geometry optimizations

To demonstrate the effect of the radius of the nanodisks (r) and the period (p), the absorption spectrum is calculated and shown for different values of r, for p = 440, 460, 480, 500, 520 and 540 nm, in Figs. 3(a), (b), (c), (d), (e), and (f), respectively. It is noteworthy that h is fixed at 180 nm, and the materials are chosen in accordance to Fig. 1. Similarly, calculations are performed for the case of varying the height of nanodisks (h), and for above-mentioned values of p at r = 140 nm. The results are shown in Figs. 4(a-f).

As shown in Fig. 3, by increasing the radius of nanodisks, the second, third, and fourth absorption maximums are shifted toward larger wavelengths. However, it comes at the cost of losing the strength of absorption at smaller wavelength range. In addition, a comparison of Figs. 3(a-f) reveals that increasing the period of the structure blue-shifts the absorption band and increases the strength of the absorption in small wavelength regime. As a compromise between the strength and width of the absorption band, the best result seems to be in Fig. 3(c) where p = 480 nm and r = 140 nm. As can be seen in Fig. 4, the increase in the height of nanodisks exerts the same effect on the absorption spectrum as the increase of r. Again, the results show that the best values for h and p are 180 and 480 nm, respectively. Therefore, Figs. 3 and 4 confirm that the optimal values for r, h, and p are 140, 180, and 480 nm, respectively.

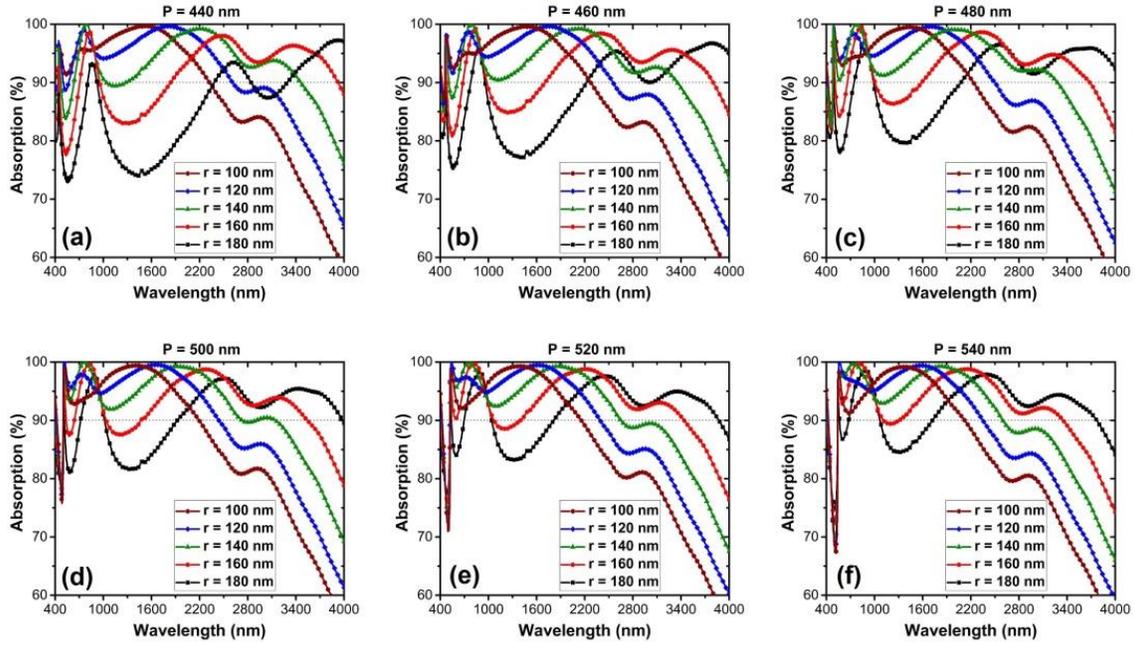

Fig. 3. Absorption spectra of the structure at different values of the radius of nanodisks (r), while period (p) is taken as (a) 440, (b) 460, (c) 480, (d) 500, (e) 520, and (f) 540 nm. In all figures, h = 180 nm and the materials are chosen in accordance to Fig. 1.

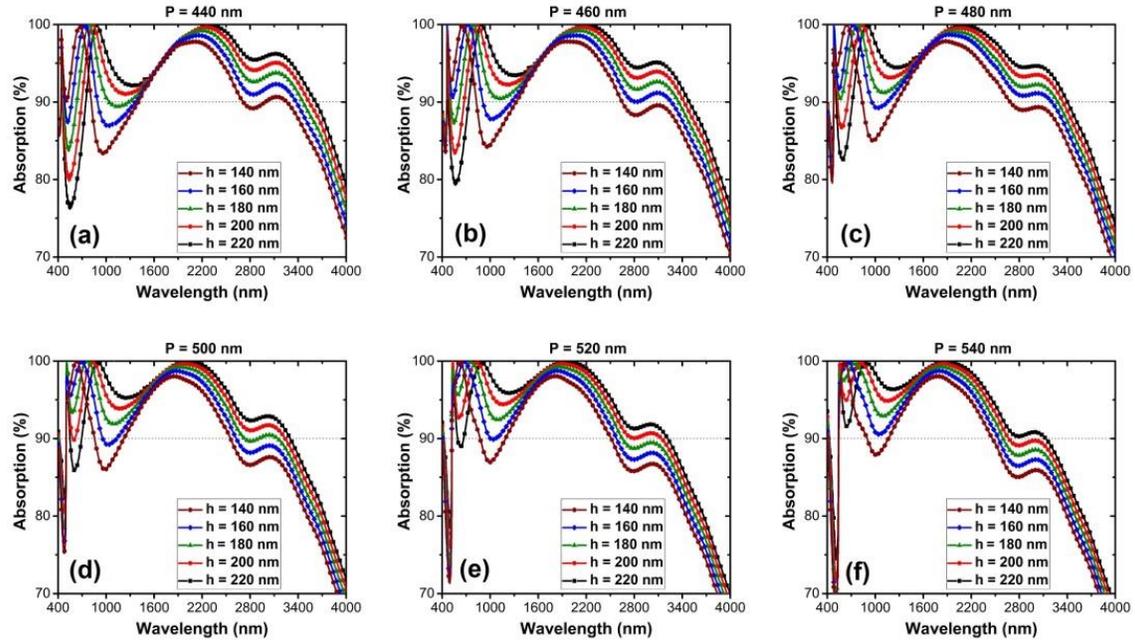

Fig. 4. Absorption spectra of the structure at different values of the height of nanodisks (h), while period (p) is taken as (a) 440, (b) 460, (c) 480, (d) 500, (e) 520, and (f) 540 nm. In all figures, r = 140 nm and the materials are chosen in accordance to Fig. 1.

## 3. Results and discussions

### 3.1 Absorption of optimal design

The absorption spectrum of the optimal structure is plotted in Fig. 5. To verify the FDTD simulations, the absorption of the structure has also been calculated using CST Microwave Studio software, which is based on the FIT. The FDTD and FIT results are shown with red solid and black dashed lines, respectively. It is shown in Fig. 5(a) that the high absorption (absorption above 90%) region starts at the visible, covers the whole NIR, and ends at the MIR regime, i.e. it is in the wavelength range extending from 478 to 3,278 nm. The bandwidth is then 2,800 nm. The relative bandwidth of absorption is calculated by RBW=BW/$\lambda_{cent}$, where BW and $\lambda_{cent}$ denote the bandwidth and the center wavelength of the absorption band, respectively. For this structure, we obtain RBW=149%. This is, to our knowledge, the best result obtained for such absorbers with one-stage-lithography that may cover a very wide wavelength range, which starts at the visible and ends within the MIR range.

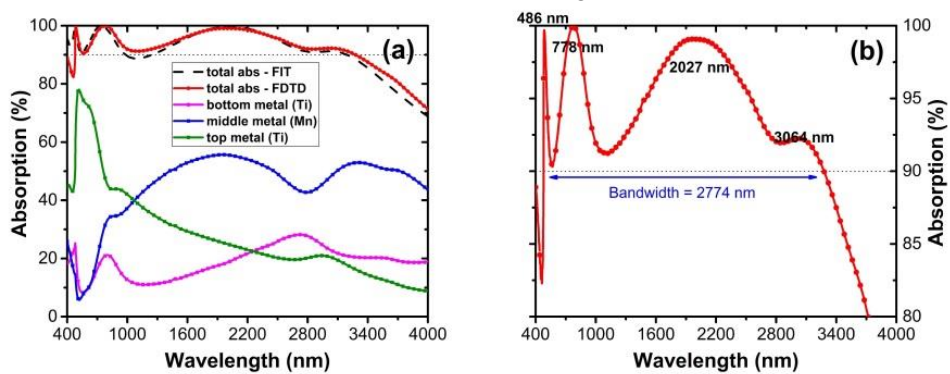

Fig. 5. (a) Absorption spectra of the optimal structure calculated by FDTD (red line) and FIT (black dashed line) methods, along with contribution of each metallic layer to the absorption of the entire structure, and (b) absorption of the optimal structure (FDTD results) with magnified ordinate axis, to clearly demonstrate the peak positions and bandwidth of the absorption.

The contribution of each metallic layer (as an absorptive component) to the resulting absorption of the structure versus the wavelength is also plotted in Fig. 5(a), where the absorption spectra of the bottom Ti layer, ultra-thin Mn layer, and top Ti nanodisk array are shown as pink, blue, and green lines, respectively.

### 3.2 Physical analysis

Now, let us get some physical insight into the observed wideband absorption. As shown in Fig. 5(b), the absorption peaks appear at 486, 778, 2,027, and 3,064 nm. The peak at the wavelength of 3,064 nm shows weaker absorption compared to the other peaks, but as it can be seen in Fig. 2(d), this peak helps the nearly perfect absorption bandwidth to increase significantly compared to the other possible choices of metals for the ultra-thin layer. Therefore, it is worth discussing the physical mechanism behind this peak, as well as the other peaks.

As the first step, we consider the absorption of each metallic layer versus the wavelength as shown in Fig. 5(a). It can be observed that at smaller wavelengths, up to around 1,000 nm, the absorption in the Ti nanodisks dominates the absorption of the Mn layer and the bottom Ti layer. This conveys that the absorption at smaller wavelengths must be mainly due to some resonance modes occurring close to the nanodisks. These resonances may lead to the trapping and absorption of light around the nanodisks, which are made of Ti, a highly lossy metal that provides the main contribution to the resulting absorption. On the other hand, at larger wavelengths, i.e. nearly after 1,000 nm, a great portion of light is absorbed by the ultra-thin, 5 nm layer of Mn, while the absorption portion of the Ti nanodisks and the thick bottom Ti layer is significantly smaller. This occurs despite the fact that the both Ti components have a much larger thickness than the Mn layer. This again signifies the crucial role of the Mn layer in the broadness of the absorption spectrum that is achieved in the studied structure.

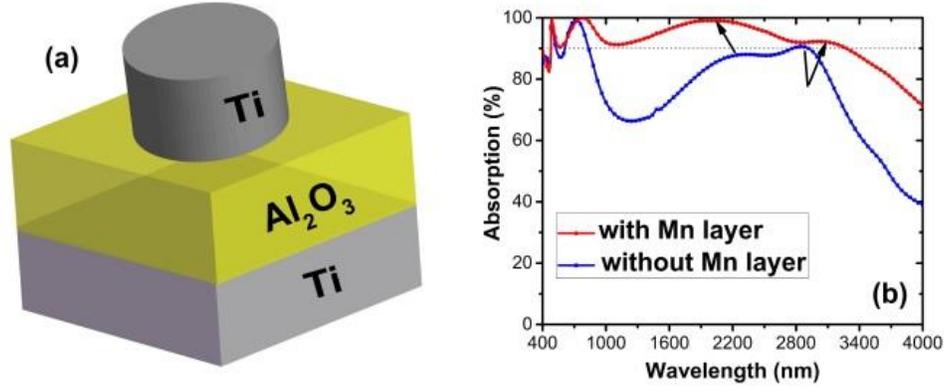

Fig. 6. (a) 3D schematic of the unit cell of the structure without the Mn layer, and (b) absorption of the optimal structure with and without the Mn layer shown in red and blue colors, respectively.

To better illustrate the significant importance of the 5 nm Mn layer, we investigate the optical response of the suggested structure without this layer, i.e. when Mn is replaced by $Al_2O_3$ and the remaining structure is a top-layer-patterned MIM configuration. The schematic of the structure without the ultra-thin Mn layer is presented in Fig. 6(a) (also see Fig. 1(d)). The absorption spectra of the structures with and without the Mn layer are shown, respectively, by red and blue lines in Fig. 6(b). The absorption band extends from 478 to 526 nm and from 610 to 836 nm for the configuration without the Mn layer, i.e. the absorption band is composed of two adjacent bands. Although the absorption spectrum has four peaks at the wavelengths of 482, 722, 2,320, and 2,848 nm in this case, the third peak at larger wavelength does not have a significant strength. Therefore, the resulting absorption band is relatively narrow (it comprises the above-mentioned adjacent bands) with the total bandwidth of 274 nm, in the case of not using the Mn layer. After adding the Mn layer, the strength and the spectral position of the first two peaks do not change significantly, but the strength of the third and fourth peaks considerably increases, and they also experience some shift that is depicted in Fig. 6(b) with the black arrows. This conveys that the first and second absorption peaks do not depend on the effect exerted by the Mn layer, while the third and fourth ones are significantly affected by the Mn layer. This is in total agreement with our observations in Figs. 5(a) and 2(d). Then by adding the Mn layer, the bandwidth gets enhanced from 274 to 2,800 nm, i.e. it experiences a tenfold increase. It is worth noting that adding an ultra-thin layer of Mn to the structure does not add considerable complexity to the fabrication process.

The absorption peaks may be due to different types of resonances such as LSP, PSP, or guided-mode resonance (GMR). The LSP resonances can occur because of the existence of nanodisks [16], the PSP mode can propagate in the continuous interface of the bottom Ti and $Al_2O_3$ layer [17], and GMR can appear in the $Al_2O_3$ layer that can work as a dielectric waveguide starting from a certain value of $d_i$ [45-47]. It is noteworthy that, as follows from the results of simulations performed for different materials of dielectric (i.e. other than $Al_2O_3$) and its thickness, GMRs may provide the main contribution to the resulting mechanism behind the first peak, in some ranges of variation of geometrical and material parameters, while PSPs do so for other sets. Moreover, it has been observed that the GMR-related and the PSP-related effects may coexist in one structure at the same frequency, and sometimes it is difficult to distinguish between the extents to which each of them contributes. For instance, based on our simulations, the dominant effect of a PSP for the first peak is obvious in case of a thinner $Al_2O_3$ layer (e.g. at $d_i$=80 nm). In such a case, the field pattern is similar to the one in [17]. At the same time, as will be exemplified below for our optimal parameters, a GMR can dominate at larger thicknesses of the dielectric layer. As the full classification of the possible physical effects and an estimation of the extent of their contribution to the resulting absorption mechanism are beyond the scope of this paper, we focus here on the optimal (main) structure.

In order to obtain some vision of the physical phenomena underlying the observed absorption peaks, the magnitude of magnetic field distribution inside the structure at the xz-plane cross section of the unit cell (see Fig. 1(c)) is simulated and presented in Fig. 7. Figs. 7(a-d) show the H-field pattern at the first, second, third, and fourth absorption peak wavelength, respectively. Because of the symmetric nature of the design, the structure is polarization insensitive for the case of normal incidence. Without any loss of generality, we may assume the electrical field vector of the normally incident wave to be in the x-direction, see Fig. 1(a). Therefore, since the propagation is in the z-direction, its magnetic field vector is in the y-direction. Field distributions in Fig. 7 show the y-component of the magnetic field ($H_y$) at the four peak wavelengths of the main structure. It is noteworthy that for this cross-section, the only component of magnetic field is $H_y$, while $H_x$ and $H_z$ are zero.

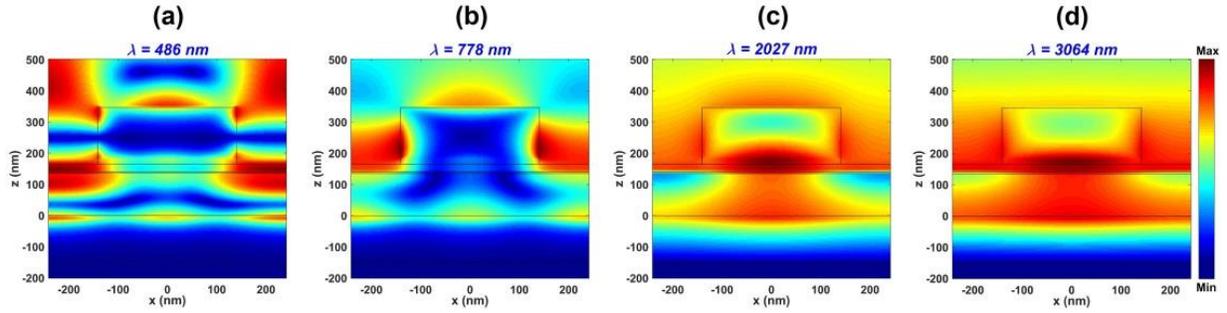

Fig. 7. The spatial distribution pattern of the magnitude of the y-component of the magnetic field ($H_y$) in the xz cross-section of the structure at the (a) first, (b) second, (c) third, and (d) fourth absorption peak

It can be observed in Fig. 7(a) that, at the first peak of the main structure, the field is strong in the $Al_2O_3$ layer that indicates possible connection to a GMR. Moreover, some features are seen, which indicate possible contribution of PSPs at the bottom $Al_2O_3$-Ti interface. At the second peak observed at the wavelength of 778 nm, the magnetic field distribution, which is shown in Fig. 7(b), indicates some localization around and between the nanodisks, so that the excitation of an LSP mode occurs here due to the coupling between the adjacent Ti disks. Field patterns of the third and fourth peaks shown in Figs. 7(c,d) are very similar to each other, and both are showing localization of the field under the nanodisks, inside the dielectric layer. The field enhancement is strong in the region above the Mn layer and under the nanodisks. For the fourth peak, the pattern demonstrates the excitation of an LSP mode (or gap-plasmon mode) inside the dielectric gap between the nanodisks and Mn, and nanodisks and bottom Ti layer [17]. However, as mentioned hereinabove, the field intensity is stronger for the LSP mode between nanodisks and the Mn layer, i.e. the gap-plasmon mode of nanodisk-Mn layer dominates. This is why the absorption of Mn layer is higher compared to the other two metallic components, at larger wavelengths (see Fig. 5(a)), and adding the Mn layer enhances the absorption of the structure at large wavelengths rather than the small-wavelength region (see Fig. 6(b)). The situation is similar but not identical for the third peak. In this case, an LSP cannot exist because the real part of permittivity is positive for all of the used materials. Therefore, we rather have here the mimicking of an LSP mode. It might be in qualitative coincidence with the recently suggested theory of surface-plasmon-like effects in all-dielectric structures [48].

To demonstrate possible effects of a GMR and an LSP mode, the absorption spectrum of a periodic array of Ti nanodisks on a 165 nm thick $Al_2O_3$ substrate is plotted in Fig. 8(a). The unit cell of the simulated structure is also shown in the inset of the figure and the dimensions are chosen as p = 480, h =

180, r = 140, $d_i$ = 165 nm. Absorption peaks are observed at 535 and 775 nm, being a reminiscence of the first and second peaks of absorption in the main structure. The peak at 775 nm almost perfectly coincides with the second absorption peak of the main structure at 778 nm. It is because they both appear due to the excitation of a coupled LSP mode between the nanodisks, and in both cases, nanodisks are the same. In turn, the first resonance at 535 nm is connected with a GMR. The reason that its spectral position does not perfectly coincide with the first peak of the main structure is that in the case of GMR, the layer below $Al_2O_3$ affects the GMR wavelength. It can be shown that adding a back-side slab below the $Al_2O_3$ layer and changing the real part of its permittivity from unity to that of Ti leads to the gradual shift of the first peak closer and closer to the first peak in the main structure (as well as to the one in the main structure but without the Mn layer). This gives us more evidence that the first peak of absorption in the main structure is connected to a GMR.

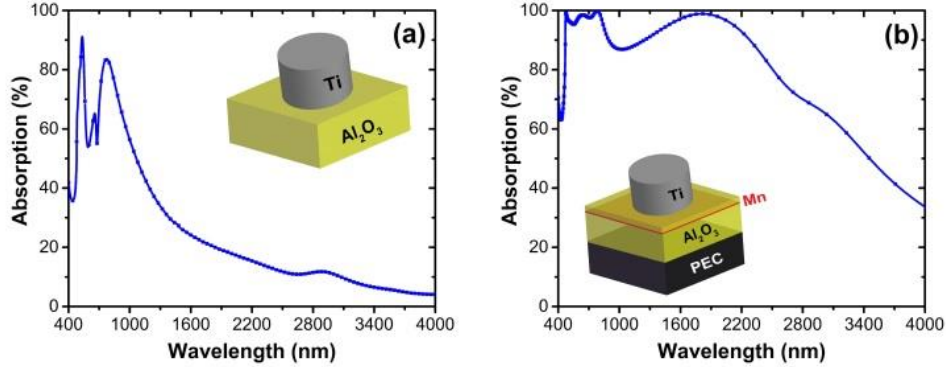

Fig. 8. (a) Absorption spectrum of the periodic array of Ti nanodisks on the $Al_2O_3$ substrate, with p = 480, h = 180, r = 140, $d_i$ = 165 nm. (b) Absorption spectrum in case when bottom Ti layer is replaced by PEC, with the same dimensions as those of optimized structure. The insets show the schematics of the simulated structures.

Next, Fig. 8(b) shows the absorption spectrum of the structure when the bottom Ti layer is replaced with a perfect electric conductor (PEC). The unit cell of the simulated structure is also shown in the inset of this figure, and the dimensions are the same as those of the optimized structure. This simulation is done to confirm the fact that the first absorption peak of the main structure is due to a GMR, and not a PSP mode. There are two absorption peaks at 477 and 783 nm, which almost perfectly match with the first and second absorption peaks of the main structure at 478 and 778 nm, respectively. This indicates that, in contrast with some of the earlier suggested absorbers [16,17], these peaks are mainly not due to the excitation of a PSP mode at the bottom $Al_2O_3$-Ti interface. Indeed, when there is a PEC at the bottom layer, any PSP mode cannot exist.

As we mentioned above, the field pattern of the first peak signifies the existence of a GMR. The phase-matching condition of the GMR in the 2D periodic structures is as follows [45,46]:

$$k_{GMR} = k_0 \sin(\theta_0) + m\left(\frac{2\pi}{p}\right) \quad (1)$$

where P is the period of the structure, $\theta_0$ is the incident angle, $K_{GMR}$ is the wavenumber of the GMR, $K_0$ is the wavenumber of the incident wave in the free space, and m is an integer number. It is noteworthy that the condition mentioned in Eq. (1) is also valid for PSPs in 2D periodic structures [16]. However, we confirmed above that the first peak of absorption for the main structure cannot be due to PSP. Eq. (1) conveys that by increasing (decreasing) the period of the structure, the spectral position of the GMR should experience a red-shift (blue-shift). To investigate this effect, the absorption spectra of the optimal structure with different values of p have been shown in Fig. 9(a). The range of wavelength is taken between 400 and 1,000 nm, for better evidence of the shifting behavior for the first two peaks. As expected, the first peak experiences a red-shift while increasing p. This confirms our earlier argument that the first peak is due to the existence of a GMR. The second peak experiences almost no shift that indicates no connection to a GMR or PSP.

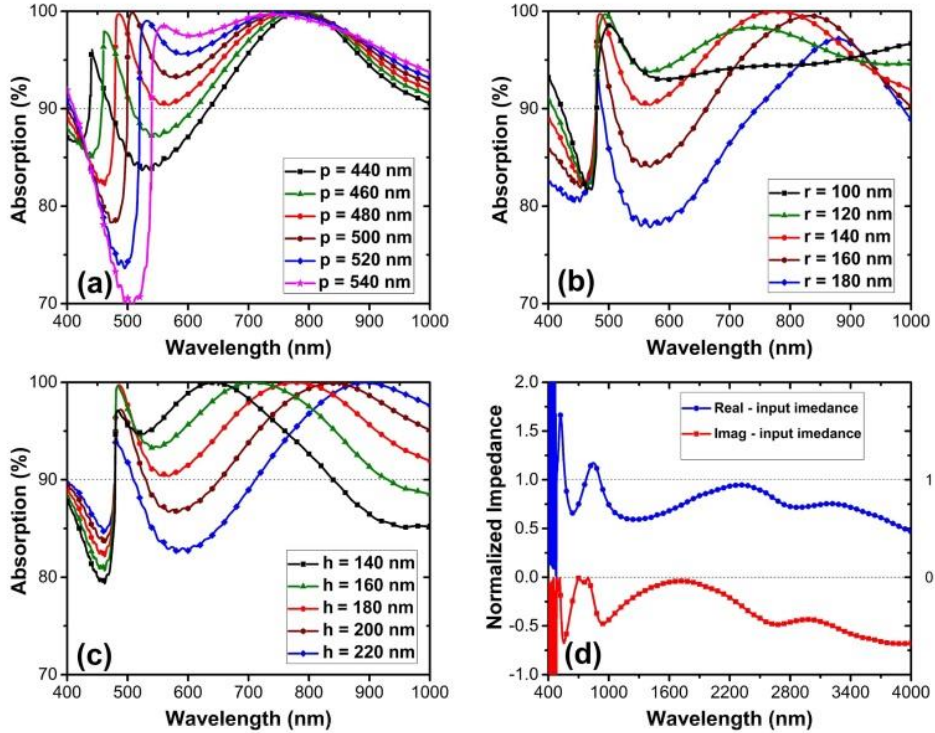

Fig. 9. Absorption spectra of the optimal structure for different values of (a) p, (b) r, and (c) h; (d) Real (blue line) and imaginary (red line) parts of the normalized input impedance of the optimal structure, Zin, calculated using S-parameters, see Eq. (2).

Let us remind here that the second peak is due to the excitation of a coupled LSP mode (Fig. 7(b)) and, fourth and third peaks are expected to appear due to the excitation of a gap-plasmon LSP mode and the mimicking of such a mode in all-dielectric case, respectively (Fig. 7(c,d)). Figures 9(b) and 9(c) demonstrate the absorption spectra of the optimal structure for different values of r and h, respectively, with the wavelength ranging from 400 to 1,000 nm. It can be observed that the increase of r or h causes an obvious red-shift in the position of the second peak, while the position of the first peak is almost insensitive to any variation of r and h. The behavior of the first peak is already expected from its origin being associated to a GMR, and not an LSP mode, while the behavior of the second peak can be clearly explained from the field pattern in Fig. 7(b). It is obvious that, because of the field localization and strong confinement of the field in the region between the adjacent disks in the case of the coupled LSP mode associated with the second peak, both r and h should affect its spectral position. Moreover, the behavior of the third and fourth peaks by changing r and h is evident in Figs. 3(c) and 4(c), respectively. It is evident in Fig. 3(c) that increasing r significantly red-shifts both the third and fourth peaks, while the increase of h exerts a very minor effect on the third peak and almost no effect on the fourth peak (Fig. 4(c)). These features can also be explained by using Figs. 7(c) and 7(d). One can see that the magnetic field for the conventional and mimicked gap-plasmon LSP modes for the both peaks are below the nanodisks, i.e. inside the dielectric ($Al_2O_3$) region and, therefore, they are associated with the radius of disks, and not with their height.

The input impedance of the entire structure, which is normalized to the free-space impedance can be calculated by using the following formula:

$$Z_{in} = \left( \frac{(1+S_{11}^2) - S_{21}^2}{(1+S_{11}^2) - S_{21}^2} \right) \tag{2}$$

where $S_{11}$ and $S_{21}$ are the elements of the S parameter's matrix, corresponding to the reflection and transmission, respectively. The results are presented in Fig. 9(d). In order to have perfect matching with free space, the real part of the normalized input impedance of the structure must be 1, and its imaginary part must be zero. At the regions where these conditions are met to some extent, i.e. even an imperfect impedance matching may lead to minimizing the reflection and, thereby, maximizing the absorption. For instance, $Z_{in}$=1.016 – j 0.029 at the second peak position, which is the cause of the strong absorption. The cause is the same for the third and fourth peaks.

After discussing the physics behind the broadband absorption, we demonstrate here why Mn is preferable for the use as the ultra-thin layer. A response to this principal question can be obtained by using the data presented in Fig. 10. The relative permittivity of Mn along with that of other metals is shown in Fig. 10. It can be observed that, in comparison with other metals, the real part of permittivity of Mn (green line) is much closer to zero in the broad wavelength range that extends from 400 to 4,000 nm. Especially, this difference is stronger at large wavelengths, i.e. above 2,000 nm. Having a small real part of permittivity leads to high penetration of the field into the material. On the other hand, again, in the range of large wavelengths above around 2,000 nm, the imaginary part of permittivity of Mn is larger compared to other metals, and this leads to stronger absorption. Both of these conditions at larger wavelengths, i.e. small real part of permittivity (high field penetration), and a large imaginary part of permittivity (strong absorption) constructively lead to the high absorption of the incident electromagnetic wave. This is the reason why Mn enables a better performance compared to other metals at relatively larger wavelengths.

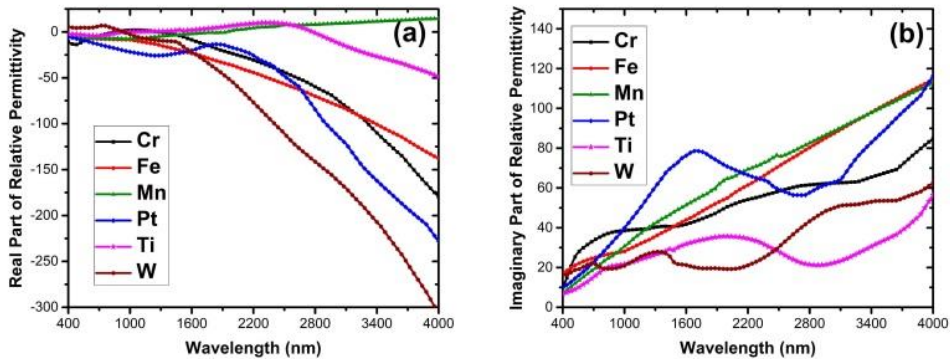

Fig. 10. (a) Real and (b) imaginary parts of the relative permittivity for Cr, Fe, Mn, Pt, Ti, and W.

### 3.3 Fabrication tolerance

Next, let us investigate the fabrication tolerance of the structure, which is an important factor for the evaluation of the prospects of mass production. As mentioned earlier, large-scale fabrication and production of the proposed structure is realizable by nanoimprint lithography, since it has only one step of lithography in its fabrication procedure.

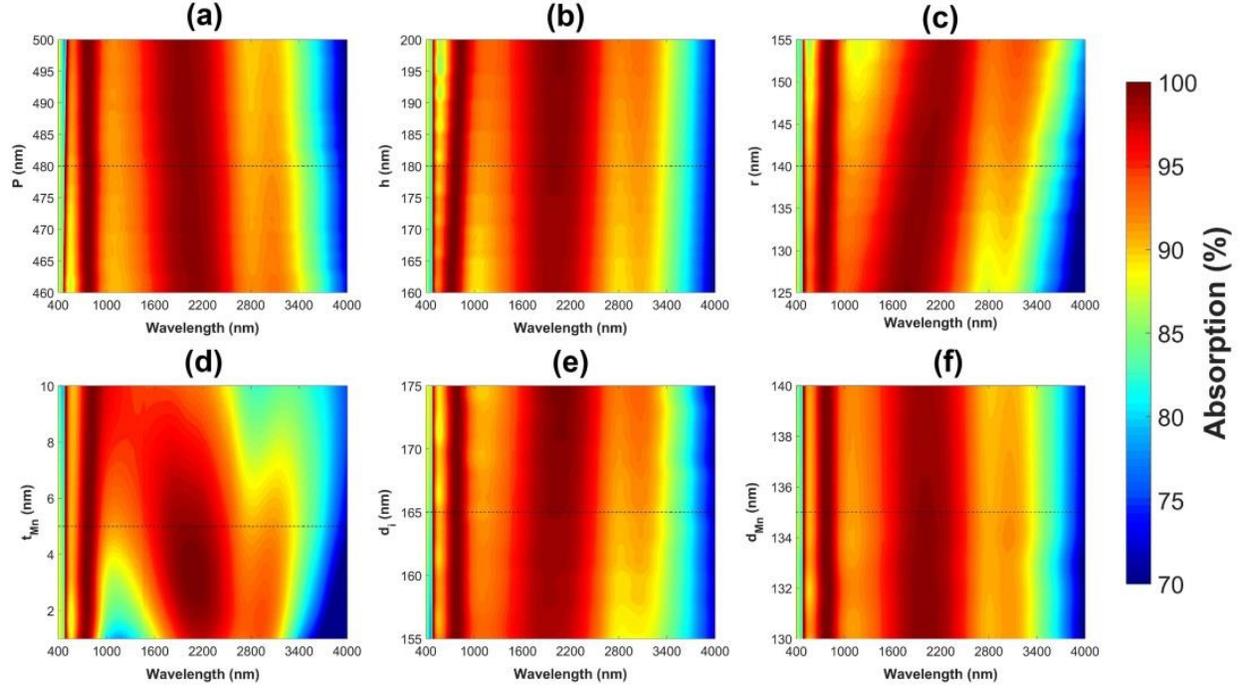

Fig. 11. Color plots of the absorption spectrum of the structure for varying the values of (a) p, (b) h, (c) r, (d) $t_{Mn}$, (e) $d_i$, and (f) $d_{Mn}$ versus the wavelength, for the purpose of the demonstration of fabrication tolerance of the optimal structure. The dashed horizontal lines indicate the optimal value of the parameter under study.

Figs. 11(a,b,c,d,e,f) show the color plots of the absorption versus wavelength for continuously varying values of p, h, r, $t_{Mn}$, $d_i$, and $d_{Mn}$, respectively, while keeping all the other parameters constant and equal to the ones of the optimized structure. These figures demonstrate the tolerance of the response of the structure to deviations from the design dimensions, which can be a result of fabrication imperfections. The optimal values of the dimensions of the studied structure are shown by the horizontal dashed black lines. The response of the structure seems to be most sensitive to the deviations of the thickness of the Mn layer ($t_{Mn}$), as shown in Fig. 11(d). In particular, one can see that deviations of $t_{Mn}$ from the selected value can lead to a band narrowing (at larger $t_{Mn}$) and band splitting (at smaller $t_{Mn}$). In addition, as can be seen in Fig. 11(c), high deviations from the optimal value of r may lead to an insignificant red-shift or blue-shift in the position of the second, third, and fourth peaks, but the general behavior of the absorption band does not get affected, significantly. The response has a very good robustness to the introduced variations in p, h, $d_i$, and $d_{Mn}$, although some relatively minor changes in the bandwidth are possible. All the plots in Fig. 11 confirm that the suggested design has a very good fabrication tolerance to possible deviations from the optimal values. This confirms that, in addition to comprising cost-effective materials, this structure is fabrication tolerant and, therefore, it is a very good candidate for future mass production.

### 3.4 Polarization and incidence angle sensitivity

As the final step, we investigate the absorption behavior of the structure for different angles of incidence for both polarizations. Figures 12(a,b) show absorption versus wavelength for incidence angles of 20, 40, and 60 degrees for TM and TE polarizations, respectively. In general, the absorption band is narrowing when the angle is increased. However, at certain values of $\lambda_{inc}$, like 40 degrees for TM polarization in Fig. 12(a), absorption is almost perfect, e.g. A>97 % at the broad range of 818< $\lambda$ <2,147 nm. The results show that the absorption remains high in a wide wavelength range for TM polarization up to incidence angles as large as 60 degrees; TE polarization is more sensitive to incident angle variations. In spite of this, rather wide bands of A>90% can exist, at least up to 40 degrees.

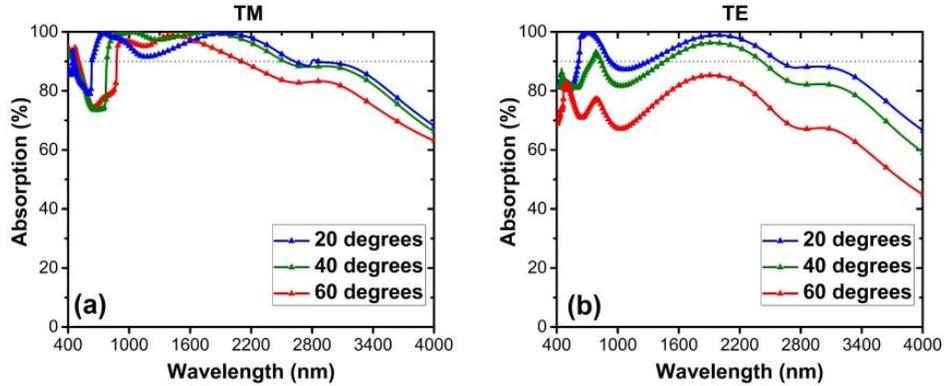

Fig. 12.  Absorption spectrum of the optimal structure for oblique incidence angles of 20, 40, and 60 degrees; calculated for the (a) TM and (b) TE polarizations.

## 4. Conclusion

An ultra-broadband, nearly perfect absorber is designed, analyzed, and characterized in detail. The studied structure is actually a top-layer-patterned MIM structure with Ti-$Al_2O_3$-Ti configuration, with an ultra-thin 5 nm layer of Mn embedded into the dielectric ($Al_2O_3$) layer. It has been shown that adding just a very thin layer of Mn enhances the bandwidth from 274 nm to 2,800 nm, i.e. it leads to a tenfold increase in the bandwidth. This allows covering the visible, NIR, and partially MIR range, while it does not need additional lithography steps and does not add much to the overall fabrication complexity of the structure. The structure has the relative absorption bandwidth of 149%, which is probably the best result obtained so far for such structures. It has been shown that the proposed structure keeps high absorption for both TM and TE polarizations for large angles. Since the structure only includes one step of lithography in its fabrication procedure, with the dimensions of lithographic patterns being over 180 nm, and with the high fabrication tolerance, its mass production can be easily realized by using nanoimprint lithography technology. Taking this fact into consideration along with the unprecedentedly ultra-broadband, nearly perfect absorption, one can conclude that the proposed absorber is very promising for photovoltaic and thermal emission applications. The use of Mn for one of the absorber components, the used design procedure and obtained numerical results can serve, along with the physical analysis, as a good starting point for design of various ultra-wideband absorbers that meet the requirements of a broad spectrum of applications, ranging from photovoltaics and solar cells to shielding and optical communication filters.

## Funding

Narodowe Centrum Nauki (NCN), Poland (DEC-2015/17/B/ST3/00118 -- Metasel); EU Horizon-2020 via Marie Sklodowska-Curie IF program (708200-ADVANTA); Turkish Academy of Sciences (TUBA); Research Office of Sharif University of Technology.

## Acknowledgments

Ekmel Ozbay acknowledges partial support from the TUBA. Amin Khavasi also acknowledges Research Office of Sharif University of Technology.